\documentclass{Interspeech}

\interspeechcameraready

\title{Voice Conversion for Likability Control via Automated Rating of Speech Synthesis Corpora}

\author[affiliation={1}]{Hitoshi}{Suda}
\author[affiliation={2,3,1}]{Shinnosuke}{Takamichi}
\author[affiliation={1}]{Satoru}{Fukayama}

\affiliation{}{National Institute of Advanced Industrial Science and Technology (AIST)}{Japan}
\affiliation{}{Keio University}{Japan}
\affiliation{}{The University of Tokyo}{Japan}
\email{suda.h@aist.go.jp, shinnosuke\_takamichi@keio.jp, s.fukayama@aist.go.jp}

\usepackage{comment}

\usepackage{amssymb}
\usepackage[capitalise,noabbrev]{cleveref}
\usepackage[nocompress,nosort]{cite}
\usepackage{multirow}
\usepackage{booktabs}
\usepackage{url}
\makeatletter
\let\MYcaption\@makecaption
\makeatother
\usepackage{subcaption}
\makeatletter
\let\@makecaption\MYcaption
\makeatother

\begin{document}

\maketitle

\begin{abstract}
Perceived voice likability plays a crucial role in various social interactions, such as partner selection and advertising. A system that provides reference likable voice samples tailored to target audiences would enable users to adjust their speaking style and voice quality, facilitating smoother communication. To this end, we propose a voice conversion method that controls the likability of input speech while preserving both speaker identity and linguistic content. To improve training data scalability, we train a likability predictor on an existing voice likability dataset and employ it to automatically annotate a large speech synthesis corpus with likability ratings. Experimental evaluations reveal a significant correlation between the predictor’s outputs and human-provided likability ratings. Subjective and objective evaluations further demonstrate that the proposed approach effectively controls voice likability while preserving both speaker identity and linguistic content.
\end{abstract}

\section{Introduction}

Speech conveys three types of information: (1) linguistic information, which is represented by sequences of discrete symbols; (2) paralinguistic information, such as speaking styles, which can be intentionally controlled by speakers; and (3) non-linguistic information, such as speaker identity, which is typically beyond their control\cite{Fujisaki1997-jw}.
Numerous studies have explored methods for controlling paralinguistic and non-linguistic features in synthesized speech.
For instance, in text-to-speech (TTS) systems, various studies have proposed methods to control the emotional tone and speaking style\cite{Wang2018-tp,Barakat2024-gc}.
Voice conversion techniques, such as speaker conversion and style transfer, have also been studied to control specific paralinguistic or non-linguistic features\cite{Stylianou2009-cx,Wang2018-tp}.
Typically, these techniques rely on reference audio signals or emotion labels to control the desired speech characteristics, whereas others employ natural language prompts\cite{Guo2023-ch}.
However, practical applications, such as designing voices for advertisements targeting distinct customer segments, require controlling speech characteristics based on more subjective criteria, such as target demographics.
The direct manipulation of paralinguistic and non-linguistic elements for subjective purposes would enable more adaptable and targeted voice designs.
Furthermore, this capability could facilitate the development of voice training applications that analyze a user's voice and generate a synthesized reference voice optimized for the speaker and tailored to specific audiences.

Voice likability is a fundamental and inherently subjective aspect of both paralinguistic and non-linguistic features of speech\cite{Weiss2020-dd}.
Several studies have demonstrated that voice likability significantly influences social outcomes, such as partner selection and leadership quality\cite{Collins2003-ao,Pavela-Banai2017-sz}.
Some previous studies have revealed a relationship between voice likability and fundamental frequency ($f_o$)\cite{Skrinda2014-jm}, while others have indicated contributions from factors such as phoneme durations and speech rate\cite{Burkhardt2011-tu}.
Analyses of a large corpus have confirmed that $f_o$ significantly impacts voice likability and indicated that additional acoustic features also contribute\cite{Suda2024-il}.
Since perceived voice likability varies among listeners\cite{Weiss2020-dd}, it is a multifaceted phenomenon that cannot be explained by a single acoustic feature.
Therefore, models that integrate multiple acoustic parameters are necessary to predict and control likability effectively.

This paper presents a voice conversion method that controls voice likability while preserving both speaker identity and linguistic content.
This method will enable users to enhance their vocal performance by generating exemplary voice samples.
However, since its development relies on subjective evaluations from multiple listeners for each training sample, the method suffers from scalability issues.
To overcome this limitation, we utilize an existing voice likability dataset\cite{Suda2024-il} to develop an automatic likability predictor.
The predictor enables the development of a likability control model without additional manual ratings, as it automatically assigns likability ratings to a speech synthesis corpus.
This paper details our approach to automatic likability prediction and voice conversion for likability control.
The paper also presents experimental results demonstrating the effectiveness of the proposed method.

\section{Automatic prediction of voice likability}\label{sec:predictor}

\begin{figure*}
    \centering
    \includegraphics{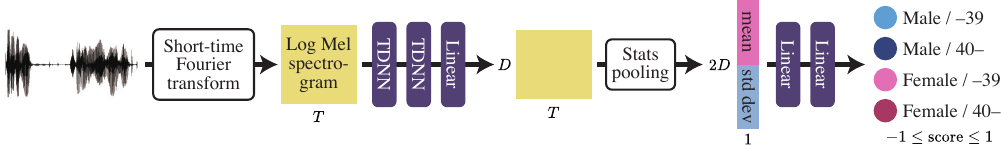}
    \caption{Architecture of the proposed voice likability predictor. The predictor outputs likability ratings ranging from $-$1 to 1 for four listener groups defined by age and gender: males under 40, males 40 or older, females under 40, and females 40 or older. $T$ denotes the number of time frames, and $D$ denotes the number of dimensions of the intermediate features.}
    \label{fig:predictor}
\end{figure*}

This section describes our approach to automatically predicting voice likability.
\Cref{fig:predictor} shows the architecture of the proposed likability predictor.
The model accepts a log-Mel spectrogram as input and employs time-delay neural networks (TDNNs) along with a statistics pooling layer, similar to the x-vector architecture\cite{Snyder2018-lk}.
The model then outputs a single time-invariant likability rating for each of four listener groups defined by gender and age.
We utilize the CocoNut-Humoresque corpus, which provides subjective ratings for 1800 voice samples along with gender and age information of the listeners\cite{Suda2024-il}.

Likability prediction is similar to mean opinion score (MOS) prediction for synthesized speech\cite{Patton2016-mv,Saeki2022-pd}.
Unlike MOS prediction, the ratings for a given speech sample can vary among listeners.
To leverage this characteristic, we partition the listeners into four groups based on gender and age: males under 40 ($n=202$), males 40 or older ($n=323$), females under 40 ($n=143$), and females 40 or older ($n=210$), where $n$ denotes the number of listeners in the corpus.
The model uses a shared network to predict the mean rating for each group.

The rating distribution in the corpus is concentrated around the center (i.e., near 0 when normalized to the range [$-$1, 1]).
To mitigate the effects of overfitting, we apply a post-filtering process that linearly transforms the initial predictions.
Specifically, on the validation set, we first compute the mean $\mu$ and variance $\sigma^2$ of the human-provided ratings, along with the corresponding mean $\hat{\mu}$ and variance $\hat{\sigma}^2$ of the predictions.
We then compute the adjusted ratings $\hat{y}'$ from the predictions $\hat{y}$ by
\begin{align}
\hat{y}' = \frac{\sigma}{\hat{\sigma}} (\hat{y} - \hat{\mu}) + \mu.
\end{align}
Since this transformation is linear, correlation-based evaluation metrics remain unchanged.

Several previous studies have also investigated automatic voice likability prediction\cite{Lu2012-nj,Eyben2013-er}.
In contrast to these methods, which rely on multiple openSMILE-based acoustic features, such as $f_o$, energy, and Mel-frequency cepstral coefficients (MFCCs), our single-network approach exhibits improved noise robustness.
Furthermore, the differentiable nature of our network facilitates its seamless integration into larger frameworks, such as TTS and voice conversion systems.

\section{Voice conversion for likability control}\label{sec:control}

\begin{figure}
    \centering
    \includegraphics{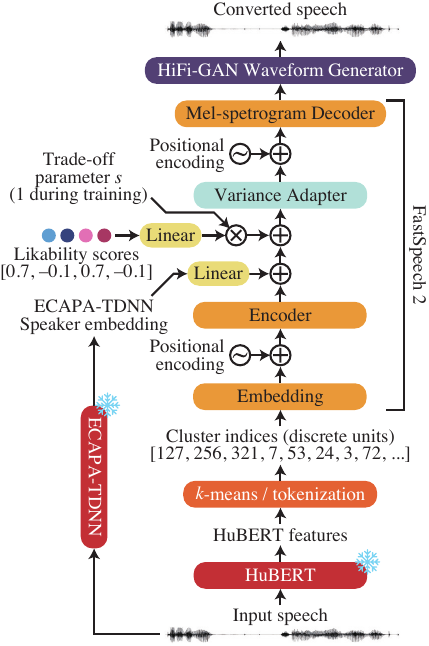}
    \caption{Architecture of the proposed voice conversion model for likability control. The model is based on FastSpeech 2\cite{Ren2021-tm} and utilizes HuBERT-based discrete units, speaker embeddings extracted with ECAPA-TDNN\cite{Desplanques2020-mx}, and target likability ratings.}
    \label{fig:converter}
\end{figure}

This section describes the proposed voice conversion method that controls voice likability while preserving both speaker identity and linguistic information.
\Cref{fig:converter} depicts the architecture of the method.
This method is based on a voice conversion approach that utilizes sequences of discrete speech units extracted from a self-supervised learning (SSL) model\cite{Huang2021-xz,van-Niekerk2022-fb}.
First, the method extracts SSL features from the training data, then applies $k$-means clustering to derive $k$ cluster centroids.
A TTS model is then trained to synthesize speech signals conditioned on three types of inputs: (1) cluster indices derived from the SSL feature sequences as discrete units; (2) speaker embeddings extracted using ECAPA-TDNN\cite{Desplanques2020-mx}; and (3) target likability ratings.
We employ hidden-unit BERT (HuBERT)\cite{Hsu2021-mm} for the SSL component and FastSpeech 2\cite{Ren2021-tm} for the TTS model.
During synthesis, sequences of consecutive identical discrete tokens are compressed into a single token.
For example, if the extracted discrete token sequence is [13, 7, 7, 21, 21, 5], the model input becomes [13, 7, 21, 5].
This strategy facilitates adjustments of speech rate based on speaker embeddings and target likability ratings.

Training this model requires a multi-speaker corpus annotated with likability ratings.
Since the CocoNut-Humoresque corpus is derived from the Coco-Nut\cite{Watanabe2023-cf}, which comprises speech samples collected from YouTube and processed via source separation, high-quality synthesis cannot be achieved using this corpus alone.
In contrast, manually rating a multi-speaker corpus is costly and is not scalable.
To overcome this limitation, we utilize likability ratings that are automatically predicted by the likability predictor described in \cref{sec:predictor}.

This approach entails a trade-off between likability control and speaker identity preservation: excessive control may compromise speaker identity, and vice versa.
To improve usability in application scenarios, our approach introduces a scalar multiplier $s$, which is applied to the embeddings of the target likability ratings.
The scalar multiplier is fixed at 1 during training but can be adjusted during inference to modulate the balance between likability control and speaker identity preservation.

\section{Experiment 1: Voice likability prediction}\label{sec:exp-prediction}

\subsection{Experimental setup}

We utilized the CocoNut-Humoresque corpus\cite{Suda2024-il} and normalized the ratings to the range [$-$1, 1].
The corpus is divided into training, validation, and test sets comprising 1500, 300, and 300 audio samples, respectively.
Each audio sample is approximately 4 seconds long, recorded in stereo at a sampling rate of \SI{44100}{Hz}.
We generated monaural signals by averaging the stereo channels and downsampling them to \SI{22050}{Hz}.

\begin{table}
    \centering
    \caption{Time contexts and the number of dimensions for each layer in the likability prediction architecture. $T_s$ denotes the total number of signal samples, $T$ denotes the total number of frames, and $t$ denotes the time-frame indices.}
    \label{tab:predictor-architecture}
    \tabcolsep = 4.5pt
    \begin{tabular}{cccc}
        \toprule
        Layer & Layer context & Input$\times$output \\
        \midrule
        spectrogram & $[0, T_s)$ & $T_s \times 80T$ \\
        frame1 & $\{t-2, t, t+2\}$ & $240 \times 32$ \\
        frame2 & $\{t-6, t-3, t, t+3, t+6\}$ & $160 \times 32$ \\
        frame3 & $[t]$ & $32 \times 32$ \\
        stats pooling & $[0, T)$ & $32T \times 64$ \\
        segment4 & $[0]$ & $64 \times 32$ \\
        segment5 & $[0]$ & $32 \times 4$ \\
        \bottomrule
    \end{tabular}
\end{table}

\Cref{tab:predictor-architecture} presents the dimensions and time contexts of each layer.
We extracted 80-bin log-Mel spectrograms up to \SI{8000}{Hz} with a hop length of 256 samples.
The loss function was the mean squared error between the predicted and human-provided ratings, and the optimal model was selected based on the Spearman rank correlation coefficient computed on the validation set.

For data augmentation, we randomly appended silence to the beginning and the end of the signals, applied reverberation, added white noise, and reversed the signals in time.
For reverberation, we used the impulse response dataset from the MIT Acoustical Reverberation Scene Statistics Survey\cite{Traer2016-cn}.

\subsection{Results}

\begin{table}
    \centering
    \caption{Performance of likability prediction. MSE and Std represent the mean squared error and the standard deviation of the test set, respectively. LCC, SRCC, and KTAU denote the linear, Spearman rank, and Kendall rank correlation coefficients, respectively. GT Std denotes the standard deviation of the human-provided (ground truth) ratings. LCC values marked with ** are statistically significant ($p < 0.01$).}
    \label{tab:predictor-results}
    \tabcolsep = 3.4pt
    \begin{tabular}{c|ccccc|c}
        \toprule
        Listeners & MSE$\downarrow$ & Std & LCC$\uparrow$ & SRCC$\uparrow$ & KTAU$\uparrow$ & GT Std \\
        \midrule
        M --39 & 0.17 & 0.40 & 0.36** & 0.39 & 0.26 & 0.32 \\
        M 40-- & 0.12 & 0.34 & 0.41** & 0.43 & 0.28 & 0.28 \\
        F --39 & 0.19 & 0.44 & 0.40** & 0.41 & 0.28 & 0.35 \\
        F 40-- & 0.17 & 0.41 & 0.38** & 0.39 & 0.26 & 0.31 \\
        \midrule
        All & 0.08 & 0.29 & 0.46** & 0.49 & 0.33 & 0.26 \\
        \bottomrule
    \end{tabular}
\end{table}

\Cref{tab:predictor-results} shows the performance of likability prediction.
A significant correlation ($r=0.46$, $p<3 \times 10^{-17}$) was observed between the predicted and human-provided ratings; therefore, the results indicate that the proposed approach effectively predicts likability ratings.
Moreover, post-filtering based on a linear transformation adjusted the standard deviation of the predicted ratings to follow that of the human-provided ratings while preserving the correlation coefficients.
Furthermore, regarding the classification of the audio samples as ``liked'' or ``disliked'' based on the mean rating, the proposed method achieved an accuracy of 74\% ($F_1 = 0.70$), demonstrating its effectiveness in predicting whether the voices were likable or not.

\section{Experiment 2: Voice likability control}

\subsection{Experimental setup}\label{sec:exp-control-setup}

We utilized two speech corpora: the JVS corpus, comprising 14997 utterances from 100 speakers\cite{Takamichi2019-tw}, and the JTES corpus, comprising 20000 utterances from 100 speakers in four emotional styles (neutral, angry, joyful, and sad)\cite{Takeishi2016-ml}.
We reserved ten neutral sentences (sentences 41--50) uttered by two female speakers (f49 and f50) and two male speakers (m49 and m50) from the JTES corpus for evaluation.
The training dataset, comprising the entire JVS corpus and 18240 utterances from the JTES corpus, was constructed to ensure that sentences or speakers did not overlap with the evaluation set.
Speaker embeddings and likability ratings were extracted for each utterance individually without averaging across speakers.

We adopted the mini-batch $k$-means algorithm\cite{Sculley2010-ys} with $k=1000$ to cluster the SSL features.
We used an open-source FastSpeech 2 implementation\footnote{\url{https://github.com/ming024/FastSpeech2}} and applied its default configurations for the LJSpeech dataset.
For waveform generation, we used a pre-trained universal HiFi-GAN model\footnote{\url{https://github.com/jik876/hifi-gan}} and fine-tuned it using Mel-spectrograms generated via teacher forcing. %
We employed the HuBERT Large model as an SSL model, which was trained on approximately 60000 hours of Japanese television broadcast audio\cite{Takizawa2025-yo}.
For speaker embedding extraction, we used an open-source ECAPA-TDNN model \texttt{speechbrain/spkrec-ecapa-voxceleb}\footnote{\url{https://huggingface.co/speechbrain/spkrec-ecapa-voxceleb}}\cite{Ravanelli2021-zm}.
The trade-off parameter $s$, described in \cref{sec:control} and \cref{fig:converter}, was set to 1 during training and to 2.5 during evaluation.

In the evaluation, we set the target likability range from $-$2 to 2.
Note that since the likability ratings in the training data were normalized to [$-$1, 1], the synthesis process extrapolates likability values when they fall outside this range.

\subsection{Objective evaluation}

\begin{figure}
    \centering
    \includegraphics[scale=0.5]{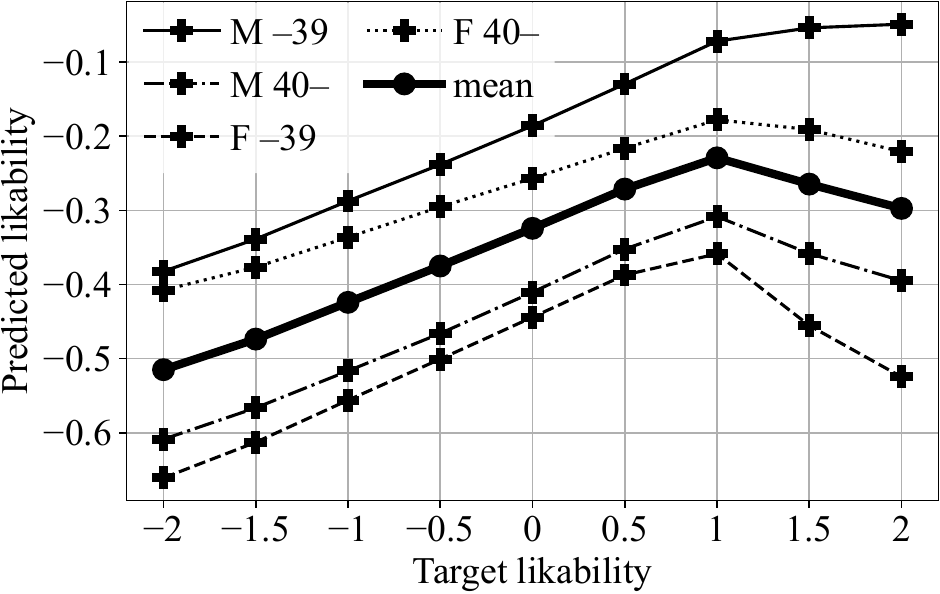}
    \caption{Predicted likability ratings of utterances with controlled likability. Thin lines represent the ratings for each gender–age listener group, while the thick line represents the average rating across all four listener groups.}
    \label{fig:likability_control}
\end{figure}

First, we evaluated the likability of the synthesized speech using the likability predictor described in \cref{sec:exp-prediction}.
\Cref{fig:likability_control} shows the results.
For all listener groups, the synthesized speech's likability was successfully controlled to follow the target ratings.
As shown in \cref{fig:likability_control}, the predicted ratings range only from $-$0.51 to $-$0.23, which is much narrower than the target range of $-$2 to 2.
Since the proposed method consistently preserves speaker identity, the model appears to adjust voice likability only within each speaker's inherent range.

\begin{figure}
    \centering
    \includegraphics[scale=0.5]{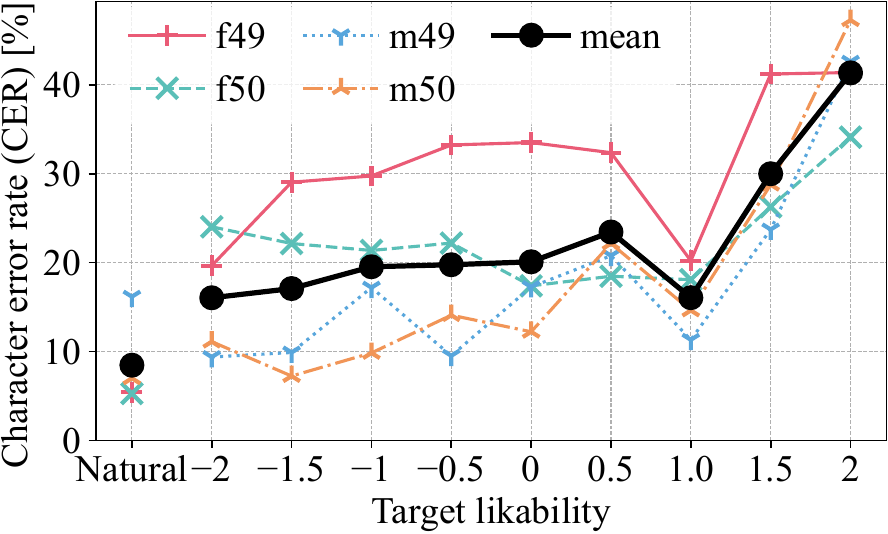}
    \caption{CER of speech recognition on the synthesized speech. The leftmost column shows the results for natural speech.}
    \label{fig:cer}
\end{figure}

Next, we evaluated the preservation of linguistic content by performing speech recognition on the synthesized speech and computing the character error rate (CER).
For speech recognition, we employed a Conformer model based on HuBERT features\cite{Gulati2020-sg,Takizawa2025-yo} trained on the LaboroTVSpeech dataset, a large-scale Japanese speech corpus\cite{Ando2021-lo}.
\Cref{fig:cer} shows the CER results.
The results indicate that the linguistic content remained consistent under likability control when the target likability was set within the range of $-$2 to 1.
However, \Cref{fig:cer} shows that the conversion process resulted in a higher CER, especially for female speakers.
This degradation indicates limitations in the voice conversion model and suggests that further improvements in clustering and model architecture are needed to achieve higher-quality conversion.

\begin{figure}
    \centering
    \includegraphics[scale=0.5]{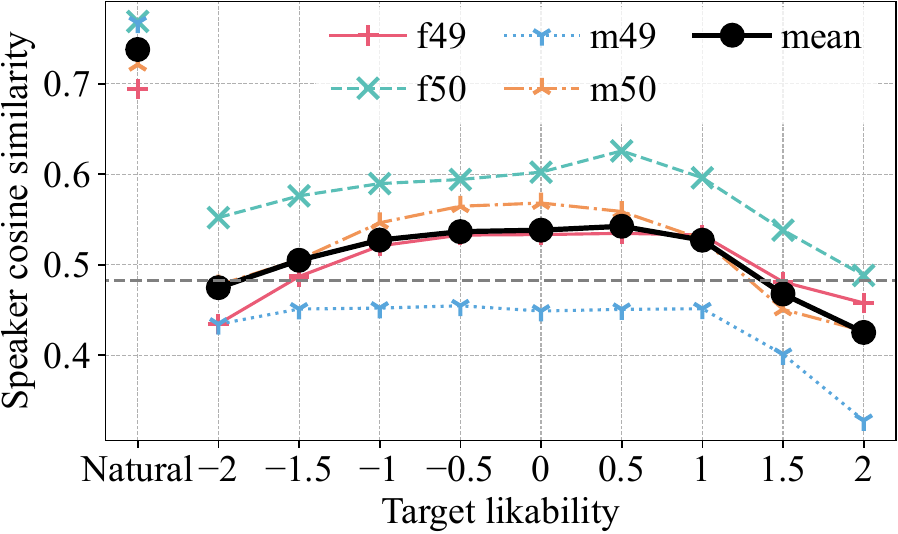}
    \caption{Cosine similarity between speaker embeddings extracted from reference and synthesized speech. The leftmost column shows the results for natural speech. The gray dashed line indicates the similarity threshold corresponding to the equal error rate (EER) computed on the JTES corpus.}
    \label{fig:similarity}
\end{figure}

Additionally, we evaluated speaker identity preservation by computing the cosine similarity between the embeddings of the reference and synthesized speech.
We used 40 neutral sentences (sentences 1–40) from the JTES corpus as a reference set and extracted embeddings with the ECAPA-TDNN model described in \cref{sec:exp-control-setup}.
The open-source implementation\footnote{\url{https://github.com/speechbrain/speechbrain}} uses a same-speaker threshold of 0.25, whereas the equal error rate (EER) condition on the JTES corpus indicates a threshold of 0.48 for speaker verification.
The results shown in \Cref{fig:similarity} indicate that speaker identity was preserved within the target ratings of $-$1 to 1, and the likability control process successfully maintained speaker identity even as target likability values varied.

\subsection{Subjective evaluation}

\begin{figure}
    \centering
    \includegraphics[trim={0 0 15pt 0},clip]{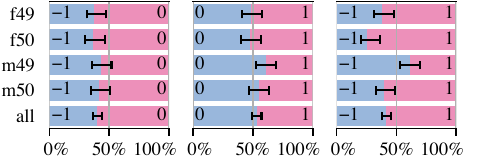}
    {\small(a) $-$1 vs. 0\hspace{31pt}(b) 0 vs. 1\hspace{31pt}(c) $-$1 vs. 1\hspace{-19pt}}
    \caption{Likability preference rates for synthesized utterances with different target likability ratings. Each row corresponds to a different speaker, while the ``all'' row represents the overall ratings. Error bars indicate the 95\% confidence intervals.} %
    \label{fig:subjective}
\end{figure}

We conducted listening tests with human participants to subjectively evaluate the effectiveness of the likability control.
We applied the likability control to 10 utterances per speaker using three target likability values: $-$1, 0, and 1.
Each listener then evaluated three randomly selected sentences per speaker for each pairing condition ($-$1 vs. 0, 0 vs. 1, and $-$1 vs. 1), resulting in 36 pairwise comparisons in total.
In the evaluation, 100 participants took part and were each paid 120 Japanese yen.

\Cref{fig:subjective} presents the results of the subjective evaluation.
The results indicate that, for three of the four speakers, the synthesized speech's likability was successfully controlled to follow the target values, with the exception of speaker m49.
Specifically, significant differences in perceived likability were observed between target values of $-$1 and 0, and an overall significant difference was found between target values of $-$1 and 1.
On the other hand, little difference in likability was observed between target values of 0 and 1, possibly due to a degradation in naturalness when the target value was 1.
Among the four speakers, speaker m49 exhibited an unexpected pattern: rather than increasing, perceived likability decreased as the target value increased from 0 to 1, and a significant degradation in likability was observed between target values of $-$1 and 1.
A possible explanation is that, for speaker m49, the synthesized speech failed to effectively preserve speaker identity, as indicated in \cref{fig:similarity}, resulting in an unnatural sound.

\section{Conclusion}
This paper presents a method for controlling voice likability for any speaker by extending a voice conversion approach that uses discrete speech units.
To improve the scalability of training data, we constructed a likability predictor based on a voice likability corpus and used it to automatically annotate speech synthesis corpora with likability ratings.
The results of both subjective and objective evaluation indicate the effectiveness of likability control as well as the preservation of speaker identity and linguistic content.
Future work includes extending our approach to a voice control application that enables users to specify desired voice design characteristics through natural language or parameters beyond likability.

\ifinterspeechfinal
\section{Acknowledgements}
This work was supported by JSPS KAKENHI Grant Numbers 21H04900, 23K20017, and 23K24895, and JST FOREST Program, Grant Number JPMJFR226V.
This research was partially supported by the AIST policy-based budget project ``R\&D on Generative AI Foundation Models for the Physical Domain.''
The authors would like to acknowledge Mr. Takizawa (AIST) for his support in the speech recognition evaluation.
\fi

\bibliographystyle{IEEEtran}
\bibliography{paperpile}

\end{document}